\documentclass[12pt]{article}
\usepackage{multicol}
\usepackage{amssymb}
\usepackage{amsfonts,amssymb,amsmath,amsthm}
\usepackage{epsfig}
\usepackage{slashed}
\usepackage{graphicx}
\usepackage{subfig}
\usepackage{mathrsfs}
\usepackage{bbm}
\usepackage{cite}
\usepackage{enumerate}
\usepackage{slashbox}
\usepackage{color}
\usepackage[colorlinks,linkcolor=red,anchorcolor=red,citecolor=green]{hyperref}
\makeatletter 

\makeatletter

\newcommand{\Rmnum}[1]{\expandafter\@slowromancap\romannumeral #1@}
\makeatother

\begin{document}
\renewcommand{\thefootnote}{\fnsymbol{footnote}}
\begin{titlepage}

\vspace{10mm}
\begin{center}
{\Large\bf On thermal molecular potential among micromolecules in charged AdS black holes}
\vspace{10mm}

{{\large Yan-Gang Miao${}^{}$\footnote{\em E-mail: miaoyg@nankai.edu.cn}
and Zhen-Ming Xu}${}^{}$\footnote{\em E-mail: xuzhenm@mail.nankai.edu.cn}

\vspace{3mm}
${}^{}${\normalsize \em School of Physics, Nankai University, Tianjin 300071, China}
}
\end{center}

\vspace{5mm}
\centerline{{\bf{Abstract}}}
\vspace{6mm}
Considering the unexpected similarity between the thermodynamic features of charged AdS black holes and that of the van der Waals fluid system, we calculate the number densities of black hole micromolecules and derive the thermodynamic scalar curvature for the small and large black holes on the co-existence curve based on the so-called Ruppeiner thermodynamic geometry. We reveal  that the microscopic feature of the small black hole perfectly matches that of the ideal anyon gas, and that the microscopic feature of the large black hole matches that of the ideal Bose gas. More importantly, we investigate the issue of molecular potential among micromolecules of charged AdS black holes, and point out explicitly that the well-known experiential Lennard-Jones potential is a feasible candidate to describe interactions among black hole micromolecules completely from a thermodynamic point of view. The behavior of the interaction force induced by the Lennard-Jones potential coincides with that of the thermodynamic scalar curvature. Both the Lennard-Jones potential and the thermodynamic scalar curvature offer a clear and reliable picture of microscopic structures for the small and large black holes on the co-existence curve for charged AdS black holes.

\vspace{5mm}
\noindent
{\bf PACS Number(s)}: 04.70.Dy, 04.70.-s, 05.70.Ce

\vspace{5mm}

\end{titlepage}

\newpage
\renewcommand{\thefootnote}{\arabic{footnote}}
\setcounter{footnote}{0}
\setcounter{page}{2}
\pagenumbering{arabic}
\tableofcontents
\vspace{1cm}

\section{Introduction}
It is widely accepted that black holes are the most probable candidate to provide a bridge between a possible quantum theory of gravity and the classical general relativity, and also that they likely represent one of the most intriguing macroscopic objects. Even more surprising is that a black hole can be mapped to an ordinary thermodynamic system based on the groundbreaking works by Hawking and Bekenstein~\cite{SH,JDB} that the black hole has temperature and entropy on an event horizon. From then on, the thermodynamics of black holes has been being an active and fascinating field of research~\cite{JMB,RW,SC}, and also being regarded as the most feasible example with semi-classical quantum gravity effects. While intriguing, the most important discovery in this active field is~\cite{SHP} the phase transition of black holes in the anti-de Sitter (AdS) spacetime. Furthermore, this result has been extended to a variety of more complicated cases, especially to the case of charged AdS black holes in which an analytical analogy with the van der Waals fluid system can be made~\cite{CEJM,DSJT,BPD,RCLY,YZ}. More precisely, a nice interplay between the thermodynamic behaviors of the charged AdS black hole and some notable features of the van der Waals fluid has been exhibited in detail. In this context, several landmarks have been derived~\cite{KM,KM1} for the charged AdS black hole, such as the $P-V$ or $T-S$ criticality, the first-order phase transition, the universal ratio of critical values, and the behaviors near critical points.

Recently, the so-called Ruppeiner thermodynamic geometry provides~\cite{GR1,GR2,GR3} phenomenologically a potential description about the types of interaction among micromolecules both in an ordinary thermodynamic system and in a black hole system. This approach thus gives a new perspective for us to study the thermodynamics of black holes, particularly in the exploration of microscopic characters of black holes. A lot of studies have shown~\cite{BMM,ABP,CC,SCWS,STS,NTW,WL,ZDSM} that the Ruppeiner thermodynamic geometry can tell us the following information through the thermodynamic scalar curvature of black holes:
\begin{itemize}
\item A positive thermodynamic scalar curvature corresponds to a repulsive interaction.
\item A negative thermodynamic scalar curvature corresponds to an attractive interaction.
\item A vanishing thermodynamic scalar curvature implies no interaction.
\end{itemize}
Besides, we have explored in our previous work~\cite{YZ1} the microscopic structures of a hairy black hole of Einstein's theory conformally coupled to a scalar field in five dimensions with the help of the Ruppeiner thermodynamic geometry, and found that the structures are similar to that of the usual ideal anyon gas, Fermi gas, and Bose gas in different parameter spaces. Therefore, it is natural to ask how to describe this interaction more intuitively and visually. Here we try to answer this question completely from a thermodynamic point of view. We take the charged AdS black hole as an example and put forward a conjecture that the well-known experiential Lennard-Jones potential is a feasible candidate which probably offers a quantitative description of the interaction force among micromolecules in the black hole. As expected, our results show that the behavior of the interaction force induced by the molecular Lennard-Jones potential coincides with that of the thermodynamic scalar curvature. Hence, the investigation of the issue of molecular potential among micromolecules can provide a new possible perspective to deeply explore internal information of black holes.

The paper is organized as follows. In section \ref{sec2}, we derive the thermodynamic scalar curvature and the number density of  micromolecules for charged AdS black holes, with which some relevant critical phenomena can be re-explained. In section \ref{sec3}, we introduce the molecular Lennard-Jones potential among micromolecules of charged AdS black holes and analyze its feasibility to depict the microscopic structure of charged AdS black holes. Finally, we devote to drawing discussions of our results in section \ref{sec4}.

\section{Critical phenomena and thermodynamic curvature}\label{sec2}
As an {\em a priori} choice, we proceed to review some basic thermodynamic critical properties of the spherically symmetric charged AdS black hole~\cite{KM}. The black hole temperature takes the following form in terms of the horizon radius $r_h$,
\begin{eqnarray}
T=\frac{1}{4\pi r_h}\left(1+\frac{3r_h^2}{l^2}-\frac{Q^2}{r_h^2}\right),\label{tem}
\end{eqnarray}
where $Q$ represents the total charge and $l$ the effective AdS curvature radius that is associated with the thermodynamic pressure
$P=3/(8\pi l^2)$. Meanwhile, the entropy $S$ and the thermodynamic volume $V$, conjugate to the temperature $T$ and the thermodynamic pressure $P$, respectively, have the forms,
\begin{equation}
S=\pi r_h^2, \qquad V=\frac{4}{3}\pi r_h^3.
\end{equation}

With the reminiscent of the van der Waals fluid, one can calculate~\cite{KM} the $P-V$ critical point or $T-S$ critical point for the charged AdS black hole at a fixed $Q$,
\begin{equation}
T_c=\frac{\sqrt{6}}{18\pi Q}, \qquad S_c=6\pi Q^2, \qquad P_c=\frac{1}{96\pi Q^2}, \qquad V_c=8\sqrt{6}\pi Q^3. \label{cv}
\end{equation}
When the temperature or the pressure is below its critical value, i.e. $T<T_c$ or $P<P_c$, there exists one first-order phase transition, called the small-large black hole phase transition. Here we regard the small black hole with thermodynamic volume $V$ below the critical value $V_c$ and the large black hole with thermodynamic volume $V$ above the critical value $V_c$. The co-existence curve of the two phases is described by the Clausius-Clapeyron equation and the Maxwell equal area law. For the sake of convenience, one usually introduces some dimensionless reduced parameters as follows,
\begin{eqnarray}
t:=\frac{T}{T_c}, \qquad s:=\frac{S}{S_c}, \qquad p:=\frac{P}{P_c}.\label{redp}
\end{eqnarray}
Note that $0\leq t\leq 1$ and $0\leq p\leq 1$. Hence, the phase diagram can be demonstrated by the equation~\cite{ESAS},
\begin{eqnarray}
t=\sqrt{\frac{p(3-\sqrt{p})}{2}},\qquad \text{or} \qquad p=\left[1-2\cos\left(\frac{{\cos}^{-1}(1-t^2)+\pi}{3}\right)\right]^2,
\end{eqnarray}
and the corresponding small and large black holes can be described by the following two equations, respectively,
\begin{eqnarray}
s_{s,l}=\frac{1}{2p}\left(\sqrt{3-\sqrt{p}}\mp \sqrt{3(1-\sqrt{p})}\right)^2, \label{sle}
\end{eqnarray}
where the subscript $s$ stands for the small black hole and $l$ for the large black hole. Note that $s_{s}$ and $s_{l}$ satisfy the inequalities, $0<s_s\leq1$ and $s_l \geq1$, respectively.

\begin{figure}
\begin{center}
\includegraphics[width=80mm]{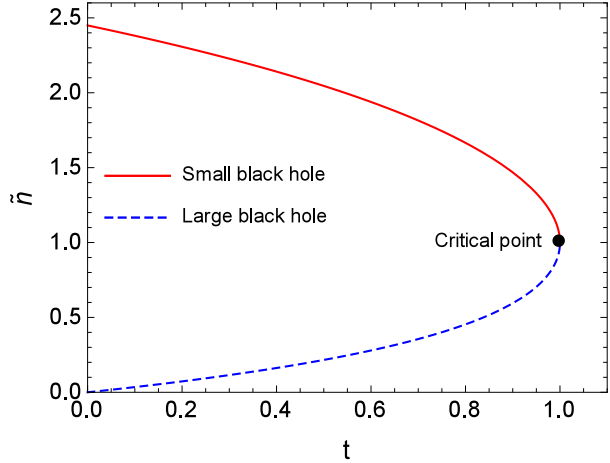}
\end{center}
\caption{The reduced number densities of the small and large black holes on the co-existence curve with respect to the reduced temperature $t$.}
\label{tu1}
\end{figure}

Another important concept is the number density of black hole micromolecules which was introduced by Wei and Liu~\cite{WL} in order to measure the microscopic degrees of freedom of the charged black hole. It is defined as
\begin{eqnarray}
n\equiv \frac{1}{2l_{\rm P}^2 r_h},\label{dn}
\end{eqnarray}
where $l_{\rm P}$ is the Planck length, $l_{\rm P}=\sqrt{\hbar G/c^3}$. For our current discussion, we need to introduce a dimensionless reduced number density $\tilde{n}\equiv n/n_c$ with the aid of the critical number density $n_c=1/(2\sqrt{6}Q)$, and thus the reduced number densities of the small and large black holes on the co-existence curve read
\begin{eqnarray}
\tilde{n}_{s,l}=\frac{\sqrt{3-\sqrt{p}}\pm \sqrt{3(1-\sqrt{p})}}{\sqrt{2}},
\end{eqnarray}
where eqs.~(\ref{sle}) and~(\ref{dn}) have been used.
The behaviors of the reduced number densities of the small and large black holes on the co-existence curve are depicted in Figure \ref{tu1}. We can see that when $t=0$, i.e. for the extremal situation, the number density is maximum for the small black hole, while it equals zero for the large black hole. With the increasing of temperature $t$, the number density decreases for the small black hole,  while it increases for the large black hole. In particular, the number density of the small black hole is equal to that of the large black hole at the critical point.

Furthermore, in order to explore the types of interaction among micromolecules of charged AdS black holes, one can appeal to the Ruppeiner thermodynamic geometry that is established on the language of Riemannian geometry. It is known that such a scheme is effective and feasible for investigating an ordinary thermodynamic system. The metric of the Ruppeiner geometry can be written in the Weinhold energy form~\cite{FW},
\begin{equation}
g_{\alpha\beta}=\frac{1}{T}\frac{\partial^2 M}{\partial X^{\alpha}\partial X^{\beta}}, \label{rg}
\end{equation}
where $X^{\alpha}$ represents an independent thermodynamic quantity, $M$ is the black hole enthalpy, and $T$ is the Hawking temperature. In our case we set coordinates $X^{\alpha}=(S,P)$ for a fixed $Q$. On the co-existence curve, according to the sign convention~\cite{GR1,GR2}, the reduced thermodynamic curvature for the small and large black holes can be written as the following forms,
\begin{eqnarray}
\tilde{R}_{s,l}\equiv \frac{R_{s,l}}{|R_c|}=-\frac{1}{2t}\cdot\frac{3s_{s,l}-1}{s_{s,l}^{5/2}},
\end{eqnarray}
where we have utilized eqs.~(\ref{tem}), ~(\ref{redp}), and~(\ref{sle}), and the critical curvature $R_c=-1/(12\pi Q^2)$. Now it is ready for us to determine the types of interaction among micromolecules of charged AdS black holes.

\begin{figure}
\begin{center}
\includegraphics[width=80mm]{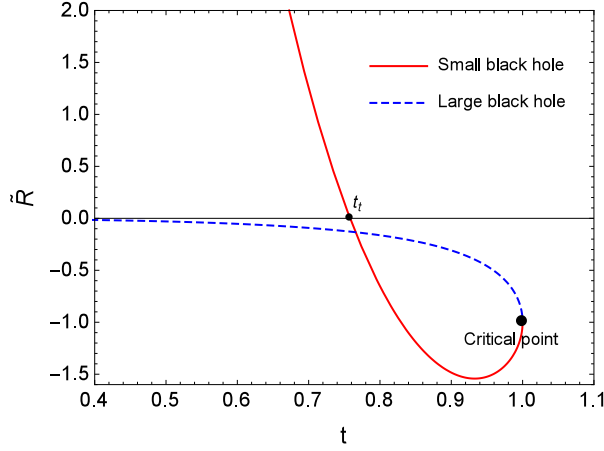}
\end{center}
\caption{The reduced thermodynamic curvature for the small and large black holes on the co-existence curve with respect to the reduced temperature $t$.}
\label{tu2}
\end{figure}

For the large black hole, we obtain $\tilde{R}_{l}<0$, impling that an attractive interaction dominates among micromolecules of charged AdS black holes and that the microscopic feature of the large black hole matches that of the ideal Bose gas~\cite{JM}. While for the small black hole, it becomes a little bit complicated. For brevity, we introduce an {\em exotic parameter sensitivity} $\gamma$ defined as
\begin{eqnarray}
\gamma\equiv\frac{Q}{(\sqrt{5}-2)l}.
\end{eqnarray}
Note that $0<\gamma \leq \gamma_c$, where the critical value $\gamma_c=1/(6\sqrt{5}-12)\simeq 0.706$. With the aid of this parameter $\gamma$, we can clearly know the information of the thermodynamic curvature for the small black hole. When $\gamma>1/2$, we get $\tilde{R}_s<0$, meaning that an attractive intermolecular interaction dominates among micromolecules of charged AdS black holes. When $\gamma<1/2$, we deduce $\tilde{R}_s>0$, implying that a repulsive intermolecular interaction dominates. The transmutation between attraction and repulsion appears at $\gamma=1/2$ and the corresponding reduced temperature and pressure read
\begin{eqnarray}
t&=&t_t\equiv \frac{3}{2}\sqrt{6(47-21\sqrt{5})}\simeq 0.758, \nonumber \\
p&=&p_t\equiv9(9-4\sqrt{5})\simeq 0.501.
\end{eqnarray}
Hence, we obtain that the microscopic feature of the small black hole matches that of the ideal anyon gas~\cite{MM,MM2}. The behaviors of the thermodynamic curvature of the small and large black holes on the co-existence curve are shown in Figure \ref{tu2}. 

We can find that for the large black hole, the thermodynamic curvature is negative. However, for the small black hole, the thermodynamic curvature can take positive, negative, and zero, and the transmutation phenomenon happens at $T_t=0.758T_c$. Moreover, according to Ruppeiner's explanation~\cite{GR2}, the microscopic degrees of freedom of black holes are carried by the Planck area pixels $l_{\rm P}^2$ and the absolute value of the thermodynamic curvature $|R|$ can be regarded as the average number of correlated pixels. Therefore, we can say that the absolute value of the thermodynamic curvature can reflect the strength of intermolecular interaction of black holes in some sense, i.e. the big absolute value of the thermodynamic curvature corresponds to strong interaction and the small absolute value of the thermodynamic curvature to weak interaction.

\section{Molecular potential}\label{sec3}
By means of the Ruppeiner thermodynamic geometry, we have a clear understanding of types of interaction among micromolecules of charged AdS black holes. For the large black hole, an attractive intermolecular interaction dominates and for the small black hole, an attractive or a repulsive intermolecular interaction exists, depending on the value of exotic parameter sensitivity. Naturally, the question is how to describe this phenomenon more intuitively. At first, it is known that the Lennard-Jones potential provides a good description of interaction among molecules in the van der Waals fluid, where the interaction force is repulsive in a short range, but attractive in a long range. Next, it is well-accepted that the thermodynamic behaviors of charged AdS black holes are very similar to that of the van der Waals fluid. As a result, it is reasonable to choose such a potential among micromolecules of charged AdS black holes which takes the form~\cite{DCJ},
\begin{eqnarray}
u(r)=4u_0\left[\left(\frac{d}{r}\right)^{12}-\left(\frac{d}{r}\right)^{6}\right],\label{mp}
\end{eqnarray}
where $r$ is center of mass separation between two molecules, $d$ is diameter of one molecule, and $u_0$ is constant. It is evident that the molecular potential vanishes, $u(r)=0$, if $r=d$. In addition, the potential takes its minimum $-u_0$ at $r=r_0\equiv 2^{1/6}d\simeq 1.122d$.

\begin{figure}
\begin{center}
\includegraphics[width=80mm]{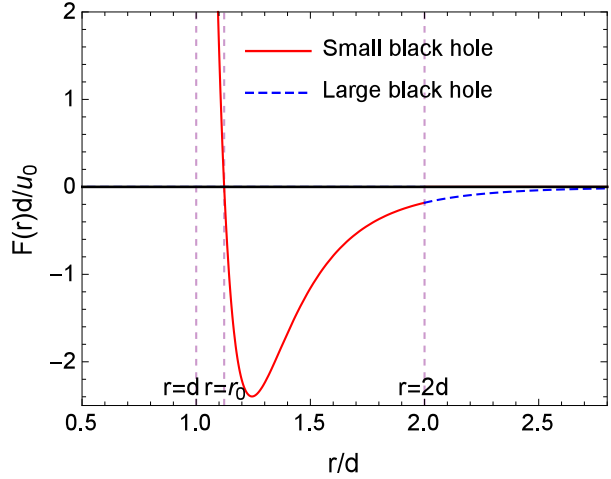}
\end{center}
\caption{The molecular interaction force $F(r)$ for the samll and large black holes on the co-existence curve of charged AdS black holes with respect to the center of mass separation $r$ in the unit of micromolecule diameter $d$.}
\label{tu3}
\end{figure}

According to the above molecular potential, the corresponding interaction force of one molecule on its neighboring molecule is
\begin{eqnarray}
F(r)\equiv -\frac{\text{d}u(r)}{\text{d}r} =\frac{24u_0}{r}\left[2\left(\frac{d}{r}\right)^{12}-\left(\frac{d}{r}\right)^{6}\right],
\end{eqnarray}
where $F(r)$ is negative (an attractive interaction) when $r>r_0$, and positive (a repulsive interaction) when $r<r_0$. Figure \ref{tu3} exhibits the behaviors of interaction force for the samll and large black holes on the co-existence curve which are summarized as follows.

\begin{itemize}
  \item The small black hole is in the {\em frozen} state when $t=0$. In this situation, the distance between two micromolecules is minimum $r=r_{\text{min}}=d$, and $F(r)$ is positive (a repulsive interaction). However, the number density is maximum as shown in Figure \ref{tu1}. The thermodynamic curvature $\tilde{R}_{s}$ is very large (see Figure \ref{tu2}), meaning that a repulsive intermolecular interaction dominates. With the increasing of temperature $t$, the reduced number density $\tilde{n}_{s}$ begins to decrease from the maximum and the distance between two micromolecules gradually increases, indicating that a repulsive force is going to weaken (see Figure \ref{tu3}). Meanwhile, the value of thermodynamic curvature $\tilde{R}_{s}$ starts to decrease. When the temperature $t$ approaches its transmutation value $t_t=0.758$, we can infer $r=r_0\simeq 1.122d$ and $F(r)=0$, which agrees with the effect of $\tilde{R}_{s}=0$. After then continuing to raise the temperature, we obtain $r>r_0$ and $F(r)<0$, implying that an attractive intermolecular interaction dominates. Correspondingly, the thermodynamic curvature $\tilde{R}_{s}$ becomes negative. In this process, we can find that the attractive force $F(r)$ has a maximum value and the absolute value of the negative $\tilde{R}_{s}$ also possesses a maximum value. Therefore, the behaviors analyzed from the molecular potential and from the thermodynamic curvature are coincident with each other.
  \item For the large black hole, the reduced number density is zero when $t=0$, implying that the distance between two micromolecules is infinity and $F(r)$ is close to zero. Meanwhile, the vanishing thermodynamic curvature $\tilde{R}_l=0$ means no interaction. With the increasing of temperature $t$, the reduced number density $\tilde{n}_l$ begins to increase from zero, indicating that the distance between two micromolecules gradually decreases from infinity, and that an attractive force $F(r)$ is getting stronger. Correspondingly,  the absolute value of the negative thermodynamic curvature $\tilde{R}_l$ is going to become large, the distance between two micromolecules is continuously decreasing but still keeping larger than $r_0$, i.e. $r>r_0$, so an attractive intermolecular interaction dominates. Thus, the behaviors of intermolecular interaction in the large black hole we obtain from the molecular potential and from the thermodynamic curvature coincide with each other.
  \item At the critical temperature $t=1$, the reduced number densities of the small and large black holes are equal to each other, that is, $\tilde{n}_s : \tilde{n}_l=1:1$. We can approximately obtain\footnote{The appearance of approximate equality is based on the asymmetry characteristics of the number densities of the small and large black holes on the co-existence curve, as shown in Figure \ref{tu1}.} $r\simeq 2d$, and conclude that an attractive intermolecular interaction dominates in the co-existence state of the small and large black holes.
\end{itemize}

Before finishing this section, we try to explain the seeming puzzle why the Coulomb potential has no effects to micromolecules of the {\em charged} AdS black hole. One reason is that the charge $Q$ is treated as a fixed external parameter but not as a thermodynamic variable due to the thermodynamical behaviors of charged AdS black holes. The other reason is that the charged AdS black hole can be set~\cite{KM} to have the same critical values (see eq.~(\ref{cv})) as that of the van der Waals fluid if an exact coincidence between the two systems is required. To this end, the relations between the black hole charge $Q$ and the parameters $a$ and $b$ in the van der Waals fluid\footnote{The equation of state for the van der Waals fluid takes the form, $k_B T=(P+a/v^2)(v-b)$, where $v$ stands for the specific volume and $k_B$ is the Boltzmann constant.} can be obtained~\cite{KM},
\begin{eqnarray}
a=\frac{3}{4\pi} \qquad \text{and} \qquad b=\frac{2\sqrt{6}Q}{3}. \label{abq}
\end{eqnarray}
On the other hand, according to the literature~\cite{DCJ}, the parameter $b$ is regarded as the effective volume (or co-volume) of a molecule and the parameter $a$ as an average value of the attractive potential energy per unit concentration. Thus, the two parameters can be given from the point of view of the Lennard-Jones potential,
\begin{eqnarray}
b= d^3, \qquad a= -2\pi \int_{r_0}^{\infty}u(r)r^2\text{d}r=\frac{20\pi d^3 u_0}{9\sqrt{2}}. \label{abdef}
\end{eqnarray}

Now we can make a possible explanation why there exists no Coulomb potential among micromolecules of black holes. As far as the relations of $Q$, $a$, and $b$ mentioned above are concerned, the charge $Q$ is just a fixed external parameter and only related to the effective volume (or co-volume) of micromolecules, which does not produce the Coulomb potential among micromolecules. As a result, it is reasonable and acceptable that the Lennard-Jones potential eq.~(\ref{mp}) describes the interaction of micromolecules for the charged AdS black hole although this potential was originally proposed to depict interactions among neutral molecules or atoms. Moreover, the coefficient $u_0$ in eq.~(\ref{mp}) can also be determined from eqs.~(\ref{abq}) and~(\ref{abdef}), $u_0=27\sqrt{3}/(160\pi^2Q)$.

\section{Summary}\label{sec4}
Based on the Ruppeiner thermodynamic geometry, we give the exact expressions of the number density and the thermodynamic scalar curvature for the small and large black holes on the co-existence curve. 
We point out explicitly that the well-known experiential Lennard-Jones potential is one of the best choices to describe interactions among micromolecules for the charged AdS black hole completely from the thermodynamic point of view. The behavior of the interaction force induced by the molecular potential coincides with that of the thermodynamic scalar curvature. Both the Lennard-Jones potential and the thermodynamic scalar curvature offer a clear picture of microscopic structures for the small and large black holes on the co-existence curve. In a sense, the Lennard-Jones potential provides a new possible perspective to deeply explore the internal information of black holes.

Finally, we make a simple discussion about our results.
\begin{itemize}
  \item The Ruppeiner thermodynamic geometry plays a bridge role in connection between the black hole thermodynamics and the molecular potential method, where the latter is first proposed in the present work. Completely from a thermodynamic point of view, an enlightening work~\cite{KM} discovered an interesting similarity between the thermodynamic features of charged AdS black holes and the critical behavior of the van der Waals fluid system, 
i.e., it is the relation between the Hawking temperature and event horizon radius that is crucial for such an analogy, for example, the  BTZ black holes behave like an ideal gas. In other words, the charged AdS black holes correspond to a system with interaction, while the  BTZ black hole is such a system with no interaction. On the other hand, in the light of Ruppeiner thermodynamic geometry, the thermodynamic curvature plays a vital role in connection between macroscopic and microscopic 
scales in a thermodynamic system. A non-vanishing thermodynamic curvature corresponds to a system with interaction, but a vanishing one corresponds to a system with no interaction. This has been confirmed by the study of a large number of statistical models~\cite{GR1}. The situation is same for black holes, for instance, the thermodynamic curvature is non-vanishing for the charged AdS black holes~\cite{SCWS,STS,NTW}, while it is vanishing for the BTZ black holes~\cite{ABP,CC}. These results are coincident with the conclusion from black hole thermodynamics. Based on the information about interaction among micromolecules in black holes, it is natural for us to introduce the molecular potential.
  \item We propose a new attempt to explore constituents of black holes according to the type of interaction. For conventional statistical models, like the ideal gas or van der Waals fluid system, its constituents are known. Therefore, the mechanism of interaction between molecules is clear, that is, the molecular forces originate from an electromagnetic interaction of electrons and atomic nuclei. While for black holes, the situation is quite different and some facts should be pointed out. At first, since a black hole can possess temperature and entropy and as Boltzmann said ``if you can heat it, it has microscopic structure'', there is no doubt that a black hole should have microscopic structure. Secondly, there does not exist hitherto a good and complete theory of quantum gravity although the most likely candidate theories, string theory and loop quantum gravity theory have achieved good results to some extent. Thus, the exploration on the microscopic structure of black holes is bound to some speculative assumptions. Thirdly, it is still unclear about the constituents of black holes. Finally, 
the Ruppeiner thermodynamic geometry phenomenologically provides the information about interaction among micromolecules both in an ordinary thermodynamic system and in a black hole system. Hence, comparing to the methods of studying the usual statistical models, we can adopt an opposite process to explore the constituents of black holes, i.e., from the type of interaction to microscopic structure. So in this sense, we can say that the microscopic feature of the small black hole perfectly matches that of the ideal anyon gas, and that the microscopic feature of the large black hole matches that of the ideal Bose gas. Meanwhile, it is justified that the molecular potential method can be used to effectively model the microscopic behaviors of black holes. This method can also be regarded as a new attempt to expand black hole thermodynamics.
  \item We give the reason that we adopt the Lennard-Jones potential to model microscopic behaviors of charged AdS black holes. Because a great similarity between the thermodynamic features of charged AdS black holes and that of the van der Waals fluid system has been well-established, the Lennard-Jones potential that was adopted in the van der Waals fluid system is now a natural choice to describe microscopic behaviors of charged AdS black holes. On the other hand, according to the analysis of the thermodynamic curvature, the charged AdS black holes can exhibit both attraction and repulsion. The Lennard-Jones potential just provides a good description of the molecular forces which are repulsive at small and attractive at large distances, respectively. Moreover, it should be emphasized that the choice of interaction potential is not unique. The Lennard-Jones potential is here selected mainly due to its simpleness and easiness to extract information of interaction.
\end{itemize}

\section*{Acknowledgments}
This work was supported in part by the National Natural Science Foundation of China under grant No.11675081. The authors would like to thank the anonymous referee for the helpful comments that improve this work greatly.

\end{document}